\documentclass[12pt]{article}
\usepackage{cite}
\usepackage[pdftex]{graphicx}
\usepackage{moreverb}
\usepackage{amssymb}                    
\usepackage{amscd}                   
\usepackage{wasysym} 
\usepackage{makeidx} 
\def\FF#1{\begin{eqnarray} #1 \end{eqnarray}}

\newcommand{\re}{{\rm e}}
\newcommand{\ri}{{\rm i}}
\newcommand{\rd}{{\rm d}}
\newcommand{\nn}{\nonumber}
\newcommand{\ah}{\ensuremath{\hat a}}
\newcommand{\ad}{\ensuremath{\hat a^\dagger}}
\newcommand{\aht}{\ensuremath{\hat a^T}}

\newcommand{\adj}{\ensuremath{\hat a^{\dagger j}}}
\newcommand{\adk}{\ensuremath{\hat a^{\dagger k}}}
\newcommand{\nh}{\ensuremath{\hat n}}
\newcommand{\ro}{\ensuremath{\hat \rho}}

\newcommand{\Ah}{\ensuremath{\hat A}}
\newcommand{\Bh}{\ensuremath{\hat B}}
\newcommand{\Ih}{\ensuremath{\hat I}}

\newcommand{\xa}{\ensuremath{\alpha}}

\newcommand{\tr}{\ensuremath{{\rm \,trace\,}}}
\newcommand{\real}{\ensuremath{{\rm \,real\,}}}

\newcommand{\fti}{\ensuremath{f}}

\newcommand{\chg}{\ensuremath{{g}}}
\newcommand{\id}{\ensuremath{{\hat 1}}}
\newcommand{\ko}{\ensuremath{{\hat K_0}}}
\newcommand{\kp}{\ensuremath{{\hat K_+}}}
\newcommand{\km}{\ensuremath{{\hat K_-}}}
\newcommand{\kpm}{\ensuremath{{\hat K_\pm}}}
\newcommand{\kN}{\ensuremath{{\hat N}}}
\newcommand{\kI}{\ensuremath{{\hat I}}}
\newcommand{\Ad}{\ensuremath{{\hat A^\dagger}}}
\newcommand{\Bd}{\ensuremath{{\hat B^\dagger}}}
\newcommand{\opluslhrim}{\ensuremath{{\,\subset\!\!\!\!\!\!+\,}}}
\begin{document}
%\jl{1}
\title{Lindblad dynamics of the damped and  forced quantum harmonic oscillator: General solution}
\author{H. J. Korsch\thanks{Electronic address: h.j.korsch@gmail.com}\\
FB Physik, Technische Universit\"at Kaiserslautern\\ D-67653 Kaiserslautern, Germany}
\maketitle
\begin{abstract}
The quantum dynamics of a damped and forced harmonic oscillator  
described by a Lindblad master equation is analyzed. The master equation
is converted into a matrix-vector representation and the resulting
non-Hermitian Schr\"odinger equation is solved by Lie-algebraic 
techniques allowing the construction of the general solution for
the density operator.
\end{abstract}
\section{Introduction}
The open system dynamics of the
 density operator $\ro(t)$, a positive operator with $\tr \ro(0)=1$,
 is described by the Lindblad master equation
\FF{&&\frac{\rd\ro}{\rd t}= -\ri\, \big[\hat H,\ro\,\big]\nn\\
   &&\qquad +
   \frac{\mu}{2}\,\big(2 \ah \ro\ad -\ad \ah\ro- \ro\ad\ah\big)
   +\frac{\nu}{2}\,\big(2 \ad \ro\ah -\ah \ad\ro- \ro\ah\ad\big).
\label{Lindblad}}
for a forced harmonic oscillator with Hamiltonian
 \FF{\hat H=\omega\big(\ad\ah+1/2\big)- \fti(t)\,\big(\ad+
 \ah\big)\label{Hamiltonian}}
with a time-dependent force $\fti(t)\in \mathbb R$,
where $\ad$ and $\ah$ are the familiar bosonic creation and
destruction operators with commutator $|\ah,\ad]=1$. 
The damping coefficients $\mu$ and $\nu$ (with $\mu>\nu \ge 0$) describe 
phenomenologically the coupling to an environment representing 
(homogeneous) loss and (inhomogeneous) pumping of particles, respectively.
In the following it will be convenient to introduce the parameters
\FF{\gamma'=(\mu+\nu)/2\ge 0 \quad \textrm{and} \quad \gamma=(\mu-\nu)/2\ge 0} 
denoted as diffusion and dissipation constants \cite{Piil06}. 
We will use units with $\hbar=1$ throughout.

The case of a force-free
harmonic oscillator has been studied in detail in \cite{Fuji12}, where
a general solution is derived, i.e.~a solution $\ro(t)$ for an arbitrary initial
condition $\ro(0)$. For the forced oscillator, a special class of solutions 
of the form
\FF{\ro(t)=b(t)\,\re^{-|\beta(t)|^2/b(t)}\,\re^{\beta(t)\ad}\,\re^{\sigma(t)\ad\ah}\,\re^{\beta^*(t)\ah}\label{rhoexp1}}
could be constructed in a recent study by the author \cite{19lindblad1}
allowing for example to describe the convergence to the limit cycle distribution
$\ro^{{\rm (\ell c)}}(t)$ for
a harmonic force.  The parameters
in (\ref{rhoexp1}) are given by $b(t)=1-u(t)$, where $u(t)=\re^{\sigma(t)}$ 
is a solution of the Riccati equation \ $\dot u =\nu-(\mu+\nu)u+\mu u^2$, which does
not depend on the parameters of the Hamiltonian and can be solved in closed form.
The force enters via $\beta(t)=b(t)\alpha(t)$, where $\alpha(t)$ is a solution of
\FF{\dot\alpha=-(\ri\omega+\gamma)\alpha+\ri f(t)\,.\label{dgl-alpha}}
Alternatively the states (\ref{rhoexp1}) can be rewritten using the 
relations \cite{Loui73}
\FF{\re^{z \ah+w\ad}=\re^{z\ah}\re^{w\ad}\re^{-zw/2}
=\re^{w\ad}\re^{z\ah}\re^{zw/2}\ , \quad
\re^{y\ah}\re^{x\ad\ah}=\re^{x\ad\ah}\re^{y\re^{x}\ah}\label{BakHaus}}
as
\FF{\ro(t)=\hat D(\alpha(t))\,\ro_{\rm th}(t)\,\hat D^\dagger(\alpha(t))\label{rhoexp2}}
in terms of the thermal distribution
\FF{\ro_{\rm th}(t)=\frac{1}{Z(t)}\,\re^{\sigma(t) \ad\ah}}
 with the partition function
\FF{Z=\tr \re^{\sigma \ad\ah}=\frac{1}{1-\re^{\sigma}}=\frac{1}{b}}
and the unitary displacement operator
\FF{\hat D(\alpha)=\re^{\alpha \ad-\alpha^*\ah} 
\quad \textrm{with}\quad |\alpha\rangle=\hat D(\alpha)|0\rangle,\label{disop}}
which generates a coherent state $|\alpha\rangle$ from the oscillator ground state.
These states are known as single mode thermal coherent states \cite{OzVo91}
describing a thermal mixed state displaced in phase space. As shown in \cite{19lindblad1} such a thermal coherent state is
form-invariant under the Lindblad evolution. 

Often the driving force is time-periodic, $f(t+T)=f(t)$, as for example in the
celebrated case of  harmonic driving, and one can look for a time-periodic
solution $\ro(t+T)=\ro(t)$. Such a Floquet-solution can be found as
a thermal coherent state (\ref{rhoexp2}) with time-periodic
coefficients. Periodicity of $u(t)$ requires $u(t)=\nu/\mu$ and therefore
$b(t)=2\gamma/\mu$ and
$\sigma(t)=\log (\nu/\mu)$. The remaining parameter, the displacement
$\alpha(t)$, must be a periodic solution $\xa^{{\rm (\ell c)}} (t)$ of the differential
equation (\ref{dgl-alpha}), as for example
\FF{\xa^{{\rm (\ell c)}} (t)=
 \frac{\fti_0}{2}\Big(
\frac{\re^{\ri \Omega t}}{\omega+\Omega-\ri\gamma}
+\frac{\re^{-\ri \Omega t}}{\omega-\Omega-\ri\gamma}\Big)}
for $f(t)=f_0\cos \Omega t$. It should be noted, of course, that this
periodic distribution agrees with the long-time limit, the limit cycle
distribution
\FF{\ro^{{\rm (\ell c)}}(t)
={\textstyle \frac{2\gamma}{\mu}}\,\re^{\log \frac{\nu}{\mu}\,\big(|\xa^{\rm (\ell c)}(t)|^2
+\nh-\xa^{\rm (\ell c)}(t)\ad-\xa^{{\rm (\ell c)}^*}\!\!(t)\,\ah\big)}
\label{ro-limit}}
as discussed in \cite{19lindblad1}.\\[-1mm]

In the present notes we will construct the general solution of the Lindblad
dynamics in three steps: First the system is transformed into a matrix-vector
representation,  a  Schr\"odinger-like picture with a non-Hermitian Hamiltonian 
$\hat{\cal H}$, which 
can also be viewed as two interacting forced harmonic oscillators with
decay. In a second step this system is solved using 
Lie-algebraic techniques and finally transformed back to the original density matrix
representation.

\section{Matrix -- vector transformation}
\label{s-vec}
Linear operators acting on Hilbert space form  a linear space and an 
element $A$ of this space can also be denoted by a ket-vector $|A\rangle$.
One can define a scalar product of two elements $A$ and $B$ of this space
as
\FF{\langle A|B\rangle=\tr A^\dagger B\,.}
The linear operators on this Liouville space are also called superoperators.
The Lindblad dynamics of the density operator can be transformed into
a vector representation using the Kronecker product (see the book 
by Steeb and Hardy \cite{Stee11} as well as \cite{AmSh15})). Here we follow
the application by Fujii \cite{Fuji12} to the force-free and time-independent
harmonic oscillator. It should be noted that a similar technique 
has also been developed by Ban \cite{Ban92} (see also \cite{Schm78}).
The result is a Schr\"odinger-like equation for the vector $|\rho(t)\rangle$ representing
the density operator $\ro(t)$, namely
\FF{\ri \,\frac{\rd |\rho\rangle}{\rd t}=\hat{\cal H}|\rho\rangle .\label{schr-rho}}
The (super)Hamiltonian $\hat{\cal H}$ is given by
\FF{\hat{\cal H}=\hat{\cal H}_0+\ri\hat{\cal G}\,,}
where $\hat{\cal H}_0$ is a representation of the Hamiltonian and $\hat{\cal G}$
of the Lindblad term. 
Following Fujii \cite{Fuji12} one can use  the tensor or 
Kronecker product $\otimes$,
a bilinear product of (quadratic) matrices satisfying for instance
\FF{(A_1\otimes B_1)(A_2\otimes B_2)=(A_1A_2)\otimes (B_1B_2)
\ ,\quad (A\otimes B)^\dagger =A^\dagger\otimes B^\dagger\,,\label{propkro}}
which implies
\FF{(A\otimes B)^n=A^n\otimes B^n }
and
\FF{\re^{A\,\otimes B}=\sum_n\frac{1}{n!}(A\otimes B)^n=\sum_n\frac{1}{n!}\,A^n\otimes B^n\,.}
If one of these matrices is equal to the unit matrix $I$ this yields
\FF{\re^{A\,\otimes I}=\re^{A}\,\otimes I\ ,\ \ \re^{I\,\otimes A}=I \otimes \re^A\,.}
Furthermore we see from
\FF{(A\otimes I)(I\otimes B)=A\otimes B=(I\otimes B)(A\otimes I)}
that $A\otimes I$ and $I\otimes B$ commute and hence
\FF{\re^{A\otimes I+I\otimes B}=\re^{A\otimes I}\re^{I\otimes B}
=(\re^{A}\otimes I)(I\otimes \re^B)=\re^A\otimes \re^B\,.\label{kronexp}}

An important application of the Kronecker product, which will be employed in the
following, is the transformation of matrix differential equations 
\FF{\frac{\rd X}{\rd t}&=&AXB \quad \Longleftrightarrow \quad 
\frac{\rd\vec x}{\rd t}=(A\otimes B^T)\,\vec x\,,\label{trans1}\\
\frac{\rd X}{\rd t}&=&AX+XB \quad \Longleftrightarrow \quad 
\frac{\rd\vec x}{\rd t}=(A\otimes I +I\otimes B^T)\,\vec x\,,\label{trans2}}
where the $B^T$ is the transpose of $B$ and
\FF{ X  =(x_{jk}) \quad \Longleftrightarrow \quad \vec x =(x_{11},x_{12},\ldots,x_{21},\ldots)^T}
as well as
\FF{A XB  \ \ \Longleftrightarrow \ \  (A\otimes B^T)\,\vec x
\ ,\quad 
AX+XB \ \ \Longleftrightarrow \ \ (A\otimes I +I\otimes B^T)\,\vec x.\label{trans3}}
\vspace*{1mm}

Applying this transformation to  the damped harmonic oscillator in eqs.~(\ref{Lindblad}) and (\ref{Hamiltonian}),
the Hamiltonian appears as 
\FF{\hat{\cal H}_0=\omega\,\big(\nh\otimes\id -\id\otimes\nh\big)
-f(\ad\otimes\id+\ah\otimes\id-\id\otimes\ad-\id\otimes\ah\big)}
and the  representation of the Lindblad term is
\FF{\hat{\cal G}= \mu \ah\otimes\ah+\nu\ad\otimes\ad-\gamma'\,
\big(\nh\otimes \id+\id\otimes\nh+\id\otimes\id\big)
+\gamma\,\id\otimes\id}
with  $\ah^T=\ad$ and  the number operator
 $\nh=\hat a^\dagger\hat a$ (compare \cite{Fuji12}). Note that
$\hat{\cal H}_0$ and $\hat{\cal G}$ are both Hermitian.
This can now be rewritten in terms of the operators
\FF{\ko=\nh\otimes\id+\id\otimes\nh+\id\otimes\id
\ ,\quad \kp=\ad\otimes\ad\ ,\quad \km=\ah\otimes\ah}
with $\ko=\ko^\dagger$ , $\km=\kp^\dagger$, which satisfy the $su(1,1)$ commutator relations
\FF{\big[\ko,\kpm\big]=\pm 2\kpm\ ,\quad \big[\kp,\km\big]=-\ko\label{Kcomm}}
as can be easily shown using (\ref{propkro}).
(Note that in \cite{Fuji12} $\ko$ is denoted by $2\hat K_3$.)
In addition we define
\FF{\kI=\id\otimes\id\ \textrm{and}\quad \kN=\nh\otimes\id-\id\otimes\nh\label{defIandN}}
(denoted by $\hat K_0$ in  \cite{Fuji12}), which commute with 
$\ko$ and $\kpm$, and
\FF{\Ah=\ah\otimes\id\ ,\quad \Bh=\id\otimes\ah}
with $\Ad=\ad\otimes\id$, $\Bd=\id\otimes\ad$ and
\FF{\kp &=&\ad\otimes\ad=(\ad\otimes\id)(\id\otimes\ad)=\Ad\Bd\\
\km &=&\ah\otimes\ah=(\ah\otimes\id)(\id\otimes\ah)=\Ah\Bh\\
\ko&=&(\ad\ah)\otimes\id+\id\otimes(\ad\ah)+\id\otimes\id\\
&=&(\ad\otimes\id)(\ah\otimes\id)+(\id\otimes\ad)(\id\otimes\ah)+\id\otimes\id
=\Ad\Ah+\Bd\Bh+\Ih\,.\nn}
The commutators can be easily determined as
\FF{&&\big[\Ah,\Bh\big]= \big[\Ah,\Bd\big]=\big[\Ad,\Bh\big]
=\big[\Ad,\Bd\big]=0\,,\\[2mm]
&&\big[\Ah,\Ad\big]= \big[\Bh,\Bd\big]=\id\otimes \id=\kI}
as well as
\FF{\big[\ko,\Ah\big]=-\Ah\ ,& \big[\kp,\Ah\big]=-\Bd,&\big[\km,\Ah\big]=0\\
\big[\ko,\Ad\big]=\Ad\ ,& \big[\kp,\Ad\big]=0\ \ \,,& \big[\km,\Ad\big]=\Bh\\
\big[\ko,\Bh\big]=-\Bh\ ,& \big[\kp,\Bh\big]=-\Ad,&\big[\km,\Bh\big]=0\\
\big[\ko,\Bd\big]=\Bd\ ,& \big[\kp,\Bd\big]=0\ \ \,,& \big[\km,\Bd\big]=\Ah
\label{ABcomm}}
and
\FF{\big[\kN,\Ah\big]=-\Ah\ ,\  \ \big[\kN,\Ad\big]=\Ad\ ,\ \ 
\big[\kN,\Bh\big]=\Bh\ ,\ \ \big[\kN,\Bd\big]=-\Bd\,.}
We then obtain
\FF{\hat{\cal H}_0&=&\omega\kN -f\,(\Ad+\Ah-\Bd-\Bh)\,,\label{kronH}\\[2mm]
\hat{\cal G}&=& \mu \km+\nu\kp-\gamma'\,\ko+\gamma\,\kI\,.\label{kronG}}
\section{Lie algebraic solution}
\label{s-lie}
In this section we construct a solution of  the Schr\"odinger-like equation (\ref{schr-rho}) for the vector $|\rho(t)\rangle$, or, preferably, for the time-evolution operator defined by
$|\rho(t)\rangle=\hat U(t)|\rho(0)\rangle$, namely
\FF{\ri \,\frac{\rd }{\rd t}\,\hat U(t)=\hat{\cal H}\, \hat U(t)}
with $\hat U(0)=\hat I$ for 
the non-Hermitian Hamiltonian $\hat{\cal H}=\hat{\cal H}_0+\ri\hat {\cal G}$
and
\FF{\hat{\cal H}_0&=&\omega\kN-f(t)\,(\Ad+\Ah -\Bd-\Bh)\,,\label{lieH}\\[2mm]
\hat{\cal G}&=&\mu \km+\nu\kp-\gamma'\ko+\gamma\kI,\label{lieG}}
where the operators $\ko$ and $\kpm$ satisfy the  $su(1,1)$ commutator relations
(\ref{Kcomm})
and $\kN$ in eq.~(\ref{defIandN}) can be written as
\FF{\kN=\Ad\Ah-\Bd\Bh.}

One observes that the algebra generated by the nine operators closes and can 
be decomposed as a semidirect
sum $R\opluslhrim S$ of the simple $su(1,1)$ algebra $S=\{\ko,\kp,\km\}$ and
the radical $R=\{\Ad,\Ah,\Bd,\Bh,\Ih,\kN\}$ (see, e.g., \cite{88lie} and references 
therein). In fact, the algebra is identical to the algebra $\tilde L_9$ 
in \cite{88lie} and one can easily modify the solution described there 
in detail for the closely related algebra $L_9$.

For the further analysis we will need the $\hat\Gamma_j$-evolved operators
$\hat\Gamma_k$ of our algebra, i.e.~the operators
 \FF{\re^{x\hat\Gamma_j}\hat\Gamma_k\re^{-x\hat\Gamma_j}
 =\hat\Gamma_k+x\,[\hat\Gamma_j,\hat\Gamma_k]+
\frac{x^2}{2!}\,\big[\hat\Gamma_j,[\hat\Gamma_j,\hat\Gamma_k\big]\,\big]+\ldots\,.}
 Here we find, for example,
 \FF{\re^{x\Ah}\Ad\re^{-x\Ah}=\Ad+x\,[\Ah,\Ad]
 +\frac{x^2}{2!}\,\big[\Ah,[\Ah,\Ad\big]\,\big]+\ldots=\Ad+x,\label{adjah}}
because of $[\Ah,\Ad]=1$, and
\FF{\re^{x\Ad}\Ah\re^{-x\Ad}=\Ah+x\,[\Ad,\Ah]+\frac{x^2}{2!}\,\big [\Ad,[\Ad,\Ah\big]\,\big]+\ldots=\Ah-x.\label{adjad}}
The same relations are also valid for $\Bh$ and $\Bd$ and some more
are listed in table \ref{tab-op} (see also the similar table II in \cite{88lie}).
For future reference we also note  the relations
\FF{&&\re^{x\kN}\re^{y\Ah}\re^{-x\kN}=\re^{y\,\re^{-x}\Ah}
\,,\ \ \ \re^{x\kN}\re^{y\Ad}\re^{-x\kN}=\re^{y\,\re^{x}\Ad},\label{tab-op-1}\\
&&\re^{x\kN}\re^{y\Bh}\re^{-x\kN}=\re^{y\,\re^{x}\Bh}
\ \,,\ \ \ \re^{x\kN}\re^{y\Bd}\re^{-x\kN}=\re^{y\,\re^{-x}\Bd},\label{tab-op-2}}
which follow from those in table \ref{tab-op} by expanding the exponential function.

\begin{table}[t]
\caption{\label{tab-op}$\hat\Gamma_j$-evolved $\hat\Gamma_k$, i.e.~the
operators \ 
 $\re^{x\hat\Gamma_j}\hat\Gamma_k\re^{-x\hat\Gamma_j}$. }
\begin{center}
\begin{tabular}{c|cccc}
$\hat\Gamma_j$  $\backslash$ $\hat\Gamma_k$ & $\Ad$ & $\Ah$ & $\Bd$ & $\Bh$ \\ \hline
\rule[0mm]{0mm}{6mm}$\ko$ & $\re^{x}\Ad$ &  $\re^{-x}\Ah$ &  $\re^{x}\Bd$ &  $\re^{-x}\Bh$ \\
\rule[0mm]{0mm}{6mm}$\kp$ & $\Ad$ &  $\Ah\!-\!x\Bd$ &  $\Bd$ &  $\Bh\!-\!x\Ad$ \\
\rule[0mm]{0mm}{6mm}$\km$ & $\Ad\!+\!x\Bh$ &  $\Ah$ &  $\Bd\!+\!x\Ah$ &  $\Bh$ \\
\rule[0mm]{0mm}{6mm}$\kN$ & $\re^{x}\Ad$ &  $\re^{-x}\Ah$ &  $\re^{-x}\Bd$ &  $\re^{x}\Bh$
\end{tabular}
\end{center}
\end{table}

Separating the Hamiltonian as $\hat{\cal H} =\hat H_S+\hat H_R$ with
\FF{\hat H_S&=&\ri \mu\km+\ri\nu\kp-\ri\gamma'\ko \ \in S\\[1mm]
\hat H_R&=&\omega \kN-f\,(\Ad+\Ah-\Bd-\Bh)+\ri\gamma \kI  \ \in R\,,}
the time-evolution operator can be factorized as
\FF{\hat U=\hat U_S\hat U_R}
where $\hat U_S$ and $\hat U_R$ are solutions of
\FF{\ri \,\frac{\rd }{\rd t}\,\hat U_S=\hat H_S \,\hat U_S
\ ,\quad \ri \,\frac{\rd }{\rd t}\,\hat U_R=
\big(\hat U_S^{-1}\hat H_R \hat U_S\big) \,\hat U_R,}
as discussed in \cite{88lie}.
Note that $\hat U_S^{-1}\hat H_R \hat U_S \in R$, because $R$ is the radical.

We first have to construct $U_S$, which can be conveniently 
obtained as the exponential product
\FF{\hat U_S(t)=\re^{d_+(t)\kp}\re^{d_0(t)\,\ko}\re^{d_-(t)\km}\,,\label{exprod}}
where the $d$-coefficients are time-dependent, satisfying three coupled
differential equations \cite{88lie}. In the 
present case, however, $H_S$ does not depend on time, and
therefore we have
\FF{\hat U_S(t)= \re^{-\ri \hat H_S\,t}=
\re^{\mu t\km +\nu t\kp-\gamma't\ko}\,.\label{Umagnus2}}
This can be 
rewritten in the product form (\ref{exprod}) using the disentangling relations
\cite{Fuji12} with parameters
\FF{&&d_+(t)=\frac{\nu}{\gamma \lambda(t)}\,\sinh (\gamma t) \ ,\ \ 
d_-(t)=\frac{\mu}{\gamma \lambda(t)}\,\sinh (\gamma t)\label{dpm}\\
&&\lambda(t)=\re^{-d_0(t)}=\cosh (\gamma t)+\frac{\gamma'}{\gamma}\,\sinh (\gamma t)\,.\label{d0}}
Note that $d_\pm$, $d_0$ and $\lambda$ are real valued.

In  the next step, one has to construct $\hat U_R(t)$. First we calculate
\FF{\hat{\cal H}'_R=\hat U_S^{-1}\hat{\cal H}_R \hat U_S =
\hat U_S^{-1}\big(\omega\kN-f\,(\Ad+\Ah -\Bd-\Bh)+\ri \gamma \kI\big)\hat U_S\,. \label{HRs}}
The first and the last term in the sum (\ref{HRs}) are trivial, $\hat U_S^{-1}\kN\hat U_S=\kN$ 
and $\hat U_S^{-1}\kI\hat U_S=\kI$,  because $\kN$ and $\kI$ commute with 
$\ko$ and $\kpm$. The other terms can be easily evaluated with help of 
table \ref{tab-op}, as for example 
\FF{\hat U_S^{-1}\Ad\hat U_S&=&\re^{-d_-\km}\re^{-d_0\,\ko}\re^{-d_+\kp}
\Ad\re^{d_+\kp}\re^{d_0\,\ko}\re^{d_-\km}\nn\\
&=&\re^{-d_-\km}\re^{-d_0\,\ko}\Ad\re^{d_0\,\ko}\re^{d_-\km}\nn\\
&=&\re^{-d_0}\re^{-d_-\km}\Ad\re^{d_-\km}=\re^{-d_0}(\Ad-d_-\Bh)}
and in the same way
\FF{\hat U_S^{-1}\Ah\,\hat U_S&=&(\re^{d_0}-d_+d_-\re^{-d_0})\Ah+d_+\re^{-d_0}\Bd,\\
\hat U_S^{-1}\Bd\,\hat U_S&=&\re^{-d_0}(\Bd-d_-\Ah),\\
\hat U_S^{-1}\Bh\,\hat U_S&=&(\re^{d_0}-d_+d_-\re^{-d_0})\Bh+d_+\re^{-d_0}\Ad,}
and therefore one obtains
\FF{\hat{\cal H}'_R&=&\omega \kN -f\,(1-d_+)\re^{-d_0}(\Ad-\Bd)\nn\\
&&\quad\quad -f\,(\re^{d_0}+(1-d_+)d_-\re^{-d_0})(\Ah-\Bh)+\ri\gamma\kI\nn\\
&=&c_1\kN+c_2\Ah+c_3\Ad+c_4\Bh+c_5\Bd+c_6\kI\,,\label{HR}}
with
\FF{&&c_1=\omega\ ,\ \ c_2=-c_4=-f(\re^{d_0}+(1-d_+)d_-\re^{-d_0})\,,\nn\\ 
&&c_3=-c_5=-f(1-d_+)\re^{-d_0}\ ,\ \ c_6=\ri\gamma\,.}
One can easily show that
\FF{\re^{d_0}+(1-d_+)d_-\re^{-d_0}=(1-d_+)\re^{-d_0}=\re^{\gamma t}\label{d0dp}}
and with
\FF{c_2=c_3=-c_4=-c_5=-f(t)\,\re^{\gamma t}=c(t) \in \mathbb R \label{c2c3c4c5}}
we obtain
\FF{\hat{\cal H}'_R=\omega\kN +c(t)\,(\Ah+\Ad-\Bh-\Bd)+\ri\gamma\Ih\,.}
For the product of exponentials
\FF{\hat U_R=\re^{g_1\kN}\re^{g_2\Ah}\re^{g_3\Ad}\re^{g_4\Bh}\re^{g_5\Bd}\re^{g_6\kI}}
we find using $\re^{x\kN}\Ah=\re^{-x}\Ah\re^{x\kN}$, $\re^{x\kN}\Ad=\re^{x}\Ad\re^{x\kN}$ 
and $\re^{x\kN}\Bh=\re^{x}\Bh\re^{x\kN}$, $\re^{x\kN}\Bd=\re^{-x}\Bd\re^{x\kN}$  (see table \ref{tab-op})  as well as  $\re^{x\Ah}\Ad=(\Ad+x)\re^{x\Ah}$ and $\re^{x\Bh}\Bd=(\Bd+x)\re^{x\Bh}$
(see eq.~(\ref{adjad}))
\FF{&&\!\frac{\rd \hat U_R}{\rd t}=
\dot g_1\kN\re^{g_1\kN}\re^{g_2\Ah}\re^{g_3\Ad}\re^{g_4\Bh}\re^{g_5\Bd}\re^{g_6\kI}
+\dot g_2\re^{g_1\kN}\Ah\re^{g_2\Ah}\re^{g_3\Ad}\re^{g_4\Bh}\re^{g_5\Bd}\re^{g_6\kI}\nn\\
&&\qquad+\dot g_3\re^{g_1\kN}\re^{g_2\Ah}\Ad\re^{g_3\Ad}\re^{g_4\Bh}\re^{g_5\Bd}\re^{g_6\kI}
+\dot g_4\re^{g_1\kN}\re^{g_2\Ah}\re^{g_3\Ad}\Bh\re^{g_4\Bh}\re^{g_5\Bd}\re^{g_6\kI}\nn\\[2mm]
&&\qquad+\dot g_5\re^{g_1\kN}\re^{g_2\Ah}\re^{g_3\Ad}\re^{g_4\Bh}\Bd\re^{g_5\Bd}\re^{g_6\kI}
+\dot g_6\re^{g_1\kN}\re^{g_2\Ah}\re^{g_3\Ad}\re^{g_4\Bh}\re^{g_5\Bd}\kI\re^{g_6\kI}\nn\\[2mm]
&&\!=\!\big(\dot g_1\kN\!+\!\dot g_2\re^{-g_1}\Ah\!+\!\dot g_3\re^{g_1}\Ad\!+\!\dot g_4\re^{-g_1}\Bh
\!+\!\dot g_5\re^{g_1}\Bd\!+\! \dot g_3g_2\!+\!\dot g_5g_4\!+\!\dot g_6\big)\hat U_R.}
Inserting this and (\ref{HR}) into 
$\ri\, \frac{\rd \hat U_R}{\rd t}=\hat{\cal H}'_R\,\hat U_R$ 
and comparing both sides, we see that that the $g_j(t)$ satisfy the differential equations
\FF{&&\dot g_1=-\ri c_1\,, \ \ \dot g_2=-\ri c_2\re^{g_1}\,, \ \ \dot g_3=-\ri c_3\re^{-g_1}
\, , \ \ \dot g_4=-\ri c_4\re^{-g_1}\,,\nn\\
&&\dot g_5=-\ri c_5\re^{g_1}\, , \ \ 
\dot g_6=-\ri (c_6-c_3g_2\re^{-g_1}-c_5g_4\re^{g_1})\,,}
with $g_j(0)=0$. These equations can be
solved by quadrature, and, using $c_1=\omega$, $c_6=\ri\gamma$ and relations (\ref{c2c3c4c5}) for the $c_2$, \ldots $c_5$,
we have
\FF{&&g_1(t)=-\ri\omega t\\
&&g_2(t)\!=\!-\!\ri\! \int_0^t\!\!c(t')\re^{\!-\!\ri\omega t'}\,\rd t' \,,\ 
g_3(t)\!=\!-\ri \!\int_0^t\!\!c(t')\re^{\ri\omega t'}\,\rd t'\\
&&g_4(t)\!=\!+\ri\! \int_0^t\!\!c(t')\re^{\ri\omega t'}\,\rd t'\,,\ 
g_5(t)\!=\!\ri \!\int_0^t\!\!c(t')\re^{\!-\!\ri\omega t'}\,\rd t'\label{g2345},\\
&&g_6(t)=\gamma t+\ri \! \int_0^t\!\!c(t')\big(g(t')\re^{-\ri\omega t'}
-g^*(t')\re^{\ri\omega t'}\big)\,\rd t'\in \mathbb R\label{g6}}
and with
\FF{\chg(t)=g_3(t)=-g_2^*(t)=-g_4(t)=g_5^*(t)\label{chi}}
we finally obtain
\FF{\hat U_R(t)=\re^{-\ri\omega t\kN}\re^{-\chg^*(t)\Ah}\re^{\chg(t)\Ad}
\re^{-\chg(t)\Bh}\re^{\chg^*(t)\Bd}\re^{g_6(t)\kI}\,.\label{exprod-UR}}
We have therefore constructed the 
time-evolution operator $\hat U(t)=\hat U_S(t)\hat U_R(t)$. 

\section{Back transformation}
\label{s-back}
In order to derive the desired solution $\ro(t)$ of the Lindblad equation, we
return from the vector to the matrix representation.
First we will consider only $\hat U_S$.  Inserting $\ko$ and $\kpm$
as well as
\FF{\re^{d_0\,\ko}=\re^{d_0\,(\nh\otimes\id+\id\otimes \nh+\id\otimes \id)}
=\re^{d_0\,\id}\otimes \re^{d_0\,\nh}\otimes\re^{d_0\,\nh}
}
according to (\ref{kronexp}) yields
\FF{U_S&=&\re^{d_+\kp}\re^{d_0\,\ko}\re^{d_-\km}
=\re^{d_+\ad\,\otimes \,\ad}\re^{d_0\,\ko}\re^{d_-\ah\,\otimes \,\ah}\nn\\
&=&\re^{d_0\id}\otimes\sum_{j,k}\frac{d_+^jd_-^k}{j!k!}(\ad\,\otimes \ad)^j
(\re^{d_0\nh}\otimes \re^{d_0\nh})(\ah\,\otimes \ah)^k.}
Rewriting the operator in this sum as
\FF{&&(\ad\,\otimes \ad)^j
(\re^{d_0\nh}\otimes \re^{d_0\nh})(\ah\,\otimes \ah)^k=
(\adj \otimes \adj)(\re^{d_0\nh}\otimes \re^{d_0\nh})(\ah^k\otimes \ah^k)\nn\\[2mm]
&&=(\adj \re^{d_0\nh}\ah^k)    
\otimes(\adj \re^{d_0\nh}\ah^k)
=(\adj \re^{d_0\nh}\ah^k)\otimes (\adk\re^{d_0\nh} \ah^j)^T}
and using  $\ad=\aht$ and $\nh^T=\nh$ leads to
\FF{U_S(t)=\re^{d_0\id}\otimes\sum_{j,k}\frac{d_+^jd_-^k}{j!k!}
(\adj \re^{d_0\nh}\ah^k)\otimes (\adk\re^{d_0\nh} \ah^j)^T\label{exprod-URS-2}}
and hence
\FF{|\rho(t)\rangle=U_S(t)|\rho(0)\rangle=\re^{d_0\id}\otimes\sum_{j,k}\frac{d_+^jd_-^k}{j!k!}
(\adj \re^{d_0\nh}\ah^k)\otimes (\adk\re^{d_0\nh} \ah^j)^T|\rho(0)\rangle.}
This vector representation can be transformed back to the original matrix space as
\FF{\ro(t)&=&\re^{d_0}\sum_{j,k}\frac{d_+^{\,j}d_-^{\,k}}{j!k!}
\adj \re^{d_0\nh}\ah^k\,\ro(0)\,\adk\re^{d_0\nh} \ah^j\nn\\
&=&\re^{d_0}\sum_{j}\frac{d_+^{\,j}}{j!}
\adj \re^{d_0\nh}\Big(\sum_{k}\frac{d_-^{\,k}}{k!}\ah^k\,\ro(0)\,\adk\Big)\re^{d_0\nh} \ah^j}
(compare eq.~(\ref{trans3})). For the force-free case we have
\FF{\hat U(t)=\re^{d_+\kp}\re^{d_0\,\ko}\re^{d_-\km}\re^{g_1\kN}\re^{g_6\kI}}
with $g_1=-\ri\omega t$ and $g_6=\gamma t$.
The operators $\kN$ and $\kI$ commute with the 
$\hat K_j$ and this can be rewritten as
$\hat U(t)=\re^{d_+\kp}\re^{\hat X}\re^{d_-\km}$
with
\FF{\hat X&=&g_6\kI+g_1\kN+d_0\ko\nn\\
&=&(d_0+g_6)(\id\otimes\id)+(d_0+g_1)(\nh\otimes\id)
(d_0-g_1)(\id\otimes\nh)\,.}
In the same way as above we obtain
\FF{U(t)=\re^{d_0+g_6}\sum_{j,k}\frac{d_+^{\,j}d_-^{\,k}}{j!k!}
(\adj \re^{(d_0+g_1)\nh}\ah^k)\otimes (\adk\re^{(d_0-g_1)\nh} \ah^j)^T\label{exprod-URS-3}}
and therefore  with $\re^{-d_0}=\lambda$, $g_1=-\ri\omega t$ and $g_6=\gamma t$
\FF{\ro(t)&=&\re^{d_0+g_6}\sum_{j,k}\frac{d_+^{\,j}d_-^{\,k}}{j!k!}
\adj \re^{(d_0+g_1)\nh}\ah^k\,\ro(0)\,\adk\re^{-g_1)\nh} \ah^j\nn\\
&=&\frac{\re^{\gamma t}}{\lambda}\sum_{j}\frac{d_+^{\,j}}{j!}
\adj \re^{(-\ri\omega t-\log \lambda)\nh}\Big(\sum_{k}\frac{d_-^{\,k}}{k!}\ah^k\,\ro(0)\,\adk\Big)
\re^{(\ri\omega t-\log \lambda)\nh} \ah^j\label{rofree}}
in agreement with the formula derived in \cite{Fuji12}.\\[0mm]

Finally we will consider the general forced harmonic oscillator with time evolution operator
\FF{\hat U(t)= U_S(t) U_R(t)\ \ \textrm{with}\ \ \ 
 U_R(t)=
\re^{g_1\kN}\re^{g_6\kI}\re^{g_2\Ah}\re^{g_3\Ad}\re^{g_4\Bh}
\re^{g_5\Bd}.}
The term $\re^{g_1\kN}\re^{g_6\kI}$ can be treated exactly as described above and the
remaining terms can be rewritten as
\FF{&&\re^{g_2\Ah}\re^{g_3\Ad}\re^{g_4\Bh}\re^{g_5\Bd}
=\re^{g_2\,\ah\otimes\id}\re^{g_3\,\ad\otimes \id}\re^{g_4\id\otimes\ah}\re^{g_5\id\otimes\ad}\nn\\
&&=\big(\re^{g_2\ah}\re^{g_3\ad}\big)\otimes \big(\re^{g_4\ah}\re^{g_5\ad}\big)
=\big(\re^{g_2\ah}\re^{g_3\ad}\big)\otimes \big(\re^{g_5\ah}\re^{g_4\ad}\big)^T\,.}
Combining this with (\ref{exprod-URS-3}), we have
\FF{\hat U(t)=\re^{d_0+g_6}\sum_{j,k}\frac{d_+^jd_-^k}{j!k!}
\big(\adj \re^{(d_0+g_1)\nh}\ah^k\re^{g_2\ah}\re^{g_3\ad}\big)
\otimes \big(\re^{g_5\ah}\re^{g_4\ad}\adk\re^{(d_0-g_1)\nh} \ah^j\big)^T\label{exprod-URS-4}}
and therefore  with $\re^{d_0}=1/\lambda$
\FF{\ro(t)&=&\re^{d_0+g_6}\sum_{j,k}\frac{d_+^{\,j}d_-^{\,k}}{j!k!}
\adj \re^{(d_0+g_1)\nh}\ah^k\,\re^{g_2\ah}\re^{g_3\ad}\ro(0)\re^{g_5\ah}\re^{g_4\ad}\,\adk\re^{(d_0-g_1)\nh} \ah^j\nn\\
&=&\frac{\re^{g_6}}{\lambda}\sum_{j}\frac{d_+^{\,j}}{j!}
\adj \re^{(d_0+g_1)\nh}\ldots\nn\\
&&\qquad\quad  \times \Big(\sum_{k}\frac{d_-^{\,k}}{k!}\ah^k\,
\re^{g_2\ah}\re^{g_3\ad}\ro(0)\re^{g_5\ah}\re^{g_4\ad}\,\adk\Big)
\re^{(d_0-g_1)\nh} \ah^j}
or, inserting the $g_j$ from eq.~(\ref{chi}) and using the Baker-Hausdorff formula
(\ref{BakHaus}), we have
\FF{\re^{g_2\ah}\re^{g_3\ad}\ro(0)\re^{g_5\ah}\re^{g_4\ad}&=&
\re^{-\chg^*\ah}\re^{\chg\ad}\ro(0)\re^{\chg^*\ah}
\re^{-\chg\ad}\nn\\
&=&\re^{-|\chg|^2}\re^{-\chg^*\ah+\chg\ad}\ro(0)\re^{\chg^*\ah-\chg\ad}}
and obtain finally the general solution
\FF{\ro(t)=\frac{\re^{\delta}}{\lambda}\sum_{j}\frac{d_+^{\,j}}{j!}
\adj \re^{(-\ri\omega t-\log \lambda)\nh}\Big(\sum_{k}\frac{d_-^{\,k}}{k!}\ah^k\,\ro_0^{\, (\rm f)}(t)
\,\adk\Big)
\re^{(\ri\omega t-\log \lambda)\nh} \ah^j\label{rofinal}}
with $\delta(t)=g_6(t)-|g(t)|^2$ and
\FF{\ro_0^{\, (\rm f)}\!(t)=\hat D(\chg(t))\ro(0)\hat D^\dagger(\chg(t)),}
where $\hat D(\chg)=\re^{\chg\ad-\chg^*\ah}$
is the displacement operator (\ref{disop}). 
Note that this equation closely resembles the general density operator
(\ref{rofree}) for the force-free system. The only difference is the normalization
factor $\re^\delta$ and the replacement of $\ro(0)$ by  $\ro_0^{\, (\rm f)}(t)$,
which describes an initial state  displaced in phase space.
One should furthermore note that $\ro(t)$ is Hermitian if $\ro(0)$ is, 
and that  eq.~(\ref{rofinal}) simplifies for special choices of the initial distribution
similar as in the force-free case \cite{Fuji12,19lindblad1}.

The final density operator (\ref{rofinal}) depends on the 
 parameters $\lambda(t)$ and
$d_\pm(t)$, which are given in eqs.~(\ref{d0}), (\ref{dpm}), just
as for the force-free system. The force $f(t)$ enters only via the parameters 
\FF{\chg(t)\!&=&\ri\! \int_0^t\!\!f(t')\re^{(\gamma+\!\ri\omega) t'}\,\rd t'\,,\\
\delta(t)&=&\gamma t+\ri \! \int_0^t\!\!f(t')\big(\chg^*(t')\re^{(\gamma
+\ri\omega) t'}-\chg(t')\re^{(\gamma-\ri\omega) t'}\big)\,\rd t'
-|\chg(t)|^2\,,\label{deltat}}
given in eqs.~(\ref{g2345}) and (\ref{g6}), where
$c(t)=-f(t)\,\re^{\gamma t}$ defined in (\ref{c2c3c4c5}) has been inserted.\\

For a harmonic driving $f(t)=f_0\cos (\Omega t)$  the integrals above can be 
evaluated in closed form as
\FF{\chg(t)&=&\frac{f_0}{2}\Big(
\frac{\re^{(\ri\omega+\ri\Omega+\gamma)t}-1}{\omega+\Omega-\ri\gamma}
+\frac{\re^{(\ri\omega-\ri\Omega+\gamma)t}-1}{\omega-\Omega-\ri\gamma}
\Big),\\[2mm]
\delta(t)&=&\gamma t +\real (J(t))-|\chg(t)|^2}
with
\FF{&&J(t)=\frac{f_0^2}{4}\Bigg(
\frac{\re^{2(\gamma-\ri\Omega)t}-1}{(\gamma+\ri\omega-\ri\Omega)(\gamma-\ri\Omega)}+\frac{\re^{2(\gamma+\ri\Omega)t}-1}{(\gamma+\ri\omega+\ri\Omega)(\gamma+\ri\Omega)}\nn\\
&&\qquad\qquad \qquad \qquad-\frac{4(\omega-\ri\gamma)}{(\omega-\ri\gamma)^2-\Omega^2}
\Big(\frac{\ri(\re^{2\gamma t}-1)}{2\gamma}+\frac{2}{f_0}\,\chg^*(t)\Big)\Bigg).
}

We will close with a brief look at the density (\ref{rofinal}) 
in the long time limit in order to extract the limit cycle distribution from
this general solution.
We observe, that in this limit the dynamics
`forgets' the initial state so that we will, for simplicity, assume $\ro(0)=|0\rangle\langle 0|$,  which leads to
\FF{\ro_0^{\, (\rm f)}\!(t)=\hat D(\chg(t))|0\rangle\langle 0|\hat D^\dagger(\chg(t))
=|g(t)\rangle\langle g(t)|.}
For such a coherent state distribution formula (\ref{rofinal}) simplifies
considerably with the result
\FF{\ro(t)=(1-d_+)\re^{|g(t)|^2\re^{-2\gamma t}}\,
\re^{-\log d_+\big(g(t)\re^{-(\gamma+\ri\omega)t}\ad+g^*(t)\re^{-(\gamma-\ri\omega)t}\ah-\nh\big)}
}
(see eq.~(4.3) in \cite{Fuji12} or eq.~(24) in \cite{19lindblad1}). For long times
we have $d_+ \rightarrow \nu/\mu$ and 
$g(t)\rightarrow \re^{(\gamma+\ri \omega)t}\,\xa^{\rm (\ell c)}(t)$ and
therefore
\FF{\ro^{{\rm (\ell c)}}(t)
={\textstyle \frac{2\gamma}{\mu}}\,\re^{\log \frac{\nu}{\mu}\,\big(|\xa^{\rm (\ell c)}(t)|^2
+\nh-\xa^{\rm (\ell c)}(t)\ad-\xa^{{\rm (\ell c)}^*}\!\!(t)\,\ah\big)}
\label{ro-limit2}}
in agreement with formula (\ref{ro-limit}) in the introduction.
%
%\section*{References}
\bibliographystyle{unsrtot}
\bibliography{abbrev,publko,paper00,rest}
%\bibliography{abbrev,publko,paper60,paper70,paper80,paper90,paper00,rest}
\end{document}